\documentclass{appolb}
\usepackage[dvipdfmx]{graphicx}
\usepackage[dvipdfmx]{color}
\usepackage{bm}
\usepackage{amssymb}

\def\zI{\mathrm{i}\hspace{0.2mm}}
\def\bs#1{\boldsymbol{#1}}
\begin{document}
\title{
QUASIFISSION DYNAMICS AND STABILITY OF SUPERHEAVY SYSTEMS\thanks{Based on discussion after a talk given by S.~Heinz entitled
\textit{``Nuclear molecule formation and time delay in collisions of nuclei with $\mathit{Z_1 + Z_2 \geq 110}$\hspace{-0.25mm}"}
presented at the XXIII Nuclear Physics Workshop ,\hspace{-0.25mm},Marie \& Pierre Curie'', Kazimierz Dolny, Poland, September 27--October 2, 2016.}
}
\author{
Kazuyuki Sekizawa$^1$ and Sophia Heinz$^{2,3}$
\address{
$^1$Faculty of Physics, Warsaw University of Technology, ulica Koszykowa 75, 00-662 Warsaw, Poland\\
$^2$GSI Helmholtzzentrum f\"ur Schwerionenforschung GmbH, 64291 Darmstadt, Germany\\
$^3$Justus-Liebig-Universit\"at Gie{\ss}en, II. Physikalisches Institut, 35392 Gie{\ss}en, Germany}
}
\maketitle
\begin{abstract}
Recent experiments revealed intriguing similarities in the $^{64}$Ni+$^{207}$Pb,
$^{132}$Xe+$^{208}$Pb, and $^{238}$U+$^{238}$U reactions at energies around
the Coulomb barrier. The experimental data indicate that for all systems substantial energy
dissipation takes place, in the first stage of the reaction, although the number of transferred
nucleons is small. On the other hand, in the second stage, a large number of nucleons are
transferred with small friction and small consumption of time. To understand the observed behavior,
various reactions were analyzed based on the microscopic time-dependent Hartree-Fock (TDHF)
theory. From a systematic analysis for $^{40,48}$Ca+$^{124}$Sn, $^{40}$Ca+$^{208}$Pb,
$^{40}$Ar+$^{208}$Pb, $^{58}$Ni+$^{208}$Pb, $^{64}$Ni+$^{238}$U, $^{136}$Xe+ $^{198}$Pt,
and $^{238}$U+$^{238}$U reactions, we find that TDHF reproduces well the measured trends.
In addition, the Balian-V\'en\'eroni variational principle is applied to head-on collisions of $^{238}$U+$^{238}$U,
and the variance of the fragment masses is compared with the experimental data, showing significant
improvement. The underlying reaction mechanisms and possible future studies are discussed.
\end{abstract}

\section{Introduction}

In heavy ion reactions at energies around the Coulomb barrier, a molecule-like nuclear
system can be formed, which is called a nuclear molecule~\cite{MO} or a dinuclear system~\cite{DNS},
after mutual capture of projectile and target nuclei. At this stage, nucleons are exchanged
actively and kinetic energy is dissipated, while the system evolves towards the equilibrium.
Because of the strong Coulomb repulsion, the system can reseparate before compound
nucleus (CN) formation (quasifission, QF), typically on the timescale of ($10^{-21}$--$10^{-20}$)~sec,
resulting in a characteristic correlation between fragment masses and scattering angles
\cite{Bock(1982),Toke(QF1985),Shen(1987),duRietz(2011),duReitz(2013)}. The QF process
significantly hinders fusion of heavy nuclei leading to, \textit{e.g.}, superheavy systems with
proton numbers well beyond $Z$\,=\,100, where the fragility of the composite systems is
reflected by the small cross sections and short lifetimes of the fusion-evaporation residues.

The study of nuclear molecule formation and evolution allows to probe the stability
of superheavy nuclear systems, also if they have proton numbers far beyond the ones of heaviest
known elements. It is revealed by exit channel characteristics like mass, charge,
angular and energy distributions. Binary reactions in the superheavy collision systems
$^{64}$Ni+$^{207}$Pb ($Z$\,=\,$Z_{\rm P}$\,+\,$Z_{\rm T}$\,=\,110)~\cite{Comas(2012)}
and $^{132}$Xe+$^{208}$Pb ($Z$\,=\,136)~\cite{Heinz(2015)} were studied at the
velocity filter SHIP at GSI. In both cases, clear signatures for the formation of long-lived
nuclear molecules which rotate by large angles of 180 degrees were observed. Even in
collisions of $^{238}$U+$^{238}$U ($Z$\,=\,184), which were investigated at the
VAMOS spectrometer at GANIL~\cite{Golabek(2010)}, a noticeably large interaction
time was deduced. A comparison of the behavior of energy dissipation, interaction times,
and deformation reveals striking similarities between these three systems. To understand
the underlying reaction mechanism is the main purpose of the present article.

To describe damped collisions of heavy nuclei, various models have been developed and
applied: \textit{e.g.}~a dynamical model based on Langevin-type equations of motion
\cite{Zagrebaev(2005),Zagrebaev(2007)1}, a dinuclear system model (DNS) \cite{Adamian(1997)1,
Adamian(1997)2,Zhu(2015),Mun(2015)}, and an improved quantum molecular dynamics model
(ImQMD) \cite{ImQMD(2002),ImQMD(2004),ImQMD(2015),ImQMD(2016)}. Among those
theoretical models, the time-dependent Hartree-Fock (TDHF) theory \cite{Negele(review),
Simenel(review)} is regarded as a microscopic one which allows to investigate nuclear structure
and reaction dynamics in a unified way from nucleonic degrees of freedom. Since a phenomenological
input is only an energy density functional (EDF), which is constructed to reproduce known
properties of finite nuclei and nuclear matter, it offers non-empirical predictions\footnote{We
note that although EDF dependence of TDHF results has not been well studied to date, and should
be studied in future, QF dynamics (orientation dependence, shell effects, contact time, etc.)
in, \textit{e.g.}, the $^{64}$Ni+$^{238}$U reaction with SLy5 in Ref.~\cite{KS_KY_Ni-U} and
in the $^{48}$Ca,$^{50}$Ti+$^{249}$Bk reactions with SLy4d in Ref.~\cite{Umar(2016)} shows
very similar features.}. It has been shown that TDHF provides a fairly good description of averaged
reaction outcomes, \textit{e.g.}, mass and charge numbers of reaction products, total kinetic
energy loss (TKEL) and scattering angle. Recent studies demonstrated that the theory provides
a reliable description also for deep-inelastic and QF processes in collisions of heavy nuclei
\cite{Umar(2016),Wakhle(QF2014)Interplay,Oberacker(2014),Washiyama(2015),Hammerton(2015),
Umar(2015)}. Here, we employ the TDHF theory to understand the experimental data.

The article is organized as follows.
In Sec.~\ref{Sec:Experiment}, we outline the experimental methods of SHIP at GSI and VAMOS at GANIL.
In Sec.~\ref{Sec:Theory}, we recall the theoretical framework of TDHF.
In Sec.~\ref{Sec:Results}, we present the experimental and theoretical results and discuss underlying reaction mechanisms.
In Sec.~\ref{Sec:Summary}, a summary and a perspective are given.

\vspace{-1.3mm}
\section{Experimental methods}\label{Sec:Experiment}

In the following we briefly summarize our experimental methods.
For detailed descriptions, we refer readers to Ref.~\cite{Hofmann(2000)}
for the velocity filter SHIP at GSI and Ref.~\cite{VAMOS} for the VAMOS
spectrometer at GANIL.

The experiments on $^{64}$Ni+$^{207}$Pb and $^{132}$Xe+$^{208}$Pb
were performed at the velocity filter SHIP at GSI. We used SHIP to separate
target-like transfer and QF products, emitted to forward angles of ($0\pm 2$)
degrees, from primary beam and background events. The reaction products which
passed SHIP were implanted in a position sensitive silicon strip detector where they
were identified by $\alpha$ decay tagging. A large region of $\alpha$ emitters was
populated in both experiments where we identified nuclei with $84 \leq Z \leq 89$.
We measured velocity spectra for each isotope by scanning stepwise the electric and
magnetic field values of SHIP and registering the yields of the identified nuclei at each
setting. The velocity spectra deliver all essential information about formation and
evolution of nuclear molecules, namely, about energy dissipation, lifetimes and
rotation as well as on the deformation of the exit channel nuclei at the scission point.

The experiments on $^{238}$U+$^{238}$U were performed at the VAMOS
spectrometer at GANIL where we measured excitation functions of binary reaction
products at five different beam energies around the Coulomb barrier. The reaction
products were detected at angles of ($35\pm 5$) degrees. VAMOS was used
in a pure quadrupole mode. The magnetic rigidity B$\rho$ was optimized for
the detection of transfer products with masses below uranium. The following
detection system was used for particle identification and trajectory reconstruction:
(i)~a secondary electron detector for time-of-flight (TOF) measurements (start signal) and to trigger the data acquisition,
(ii)~two drift chambers for determining the positions ($x$,~$y$) and scattering angles,
(iii)~an ionization chamber to measure the energy loss~${\Delta}E$,
(iv)~a~500\,$\mu$m thick Si wall to measure the residual energy and for TOF measurement (stop signal).
From these parameters we obtained the mass number $A$ and the proton number $Z$ of the reaction products.
The resolutions of $A$ and $Z$ were $\Delta A/A$$\,=\,$2\% (FWHM) and $\Delta Z/Z$$\,=\,$6\% (FWHM).
The lowest accessible cross sections were about 1\,$\mu$b.

\section{Theoretical framework}\label{Sec:Theory}

In this section, we briefly recall the theoretical framework of the TDHF theory
\cite{Negele(review),Simenel(review)}. In TDHF, the many-body wave function
of the system is expressed as a single Slater determinant for all times,
\begin{equation}
\Phi(\bs{r}_1\sigma_1,\dots,\bs{r}_N\sigma_N,t)=\frac{1}{\sqrt{N!}}\det\bigl\{\phi_i(\bs{r}_j\sigma_jq_j,t)\bigr\},
\end{equation}
where $N$ ($=N_{\rm P}+N_{\rm T}$) is the total number of nucleons
in the system, and $\phi_i(\bs{r}\sigma q,t)$ ($i=1,\dots,N$) denotes the
single-particle orbitals of the $i$th nucleon. $\bs{r}$, $\sigma$, and $q$ are
spatial, spin, and isospin coordinates, respectively. The Pauli exclusion principle
is thus ensured during the entire time evolution. The time evolution of the
single-particle orbitals is governed by the TDHF equations,
\begin{equation}
\zI\hbar\frac{\partial\phi_i(\bs{r}\sigma q,t)}{\partial t} = \hat{h}(t)\phi_i(\bs{r}\sigma q,t),
\label{Eq:TDHF}
\end{equation}
where $\hat{h}(t)$ is the single-particle Hamiltonian which is dependent on
single-particle orbitals at each time through various densities and is derived
from appropriate functional derivatives of an EDF. The initial state for TDHF
calculations is taken as a product of Slater determinants for Hartree-Fock (HF)
ground states of projectile and target nuclei boosted with a relative velocity.
The relative velocity is evaluated assuming the Rutherford trajectory. By solving
the TDHF equations (\ref{Eq:TDHF}) with this initial wave function, the whole
reaction dynamics, \textit{e.g.}, energy dissipation, nucleon transfer, neck formation,
QF or fusion, is described in real-space and real-time, from nucleonic degrees of freedom.
We note that in TDHF calculations for heavy-ion reactions outlined above, neither
adjustable parameters nor empirical assumptions on the dynamics are introduced.

\vspace{-1mm}
\section{Results and discussion}\label{Sec:Results}

\begin{figure}[t]
\centerline{%
\includegraphics[height=16.6cm]{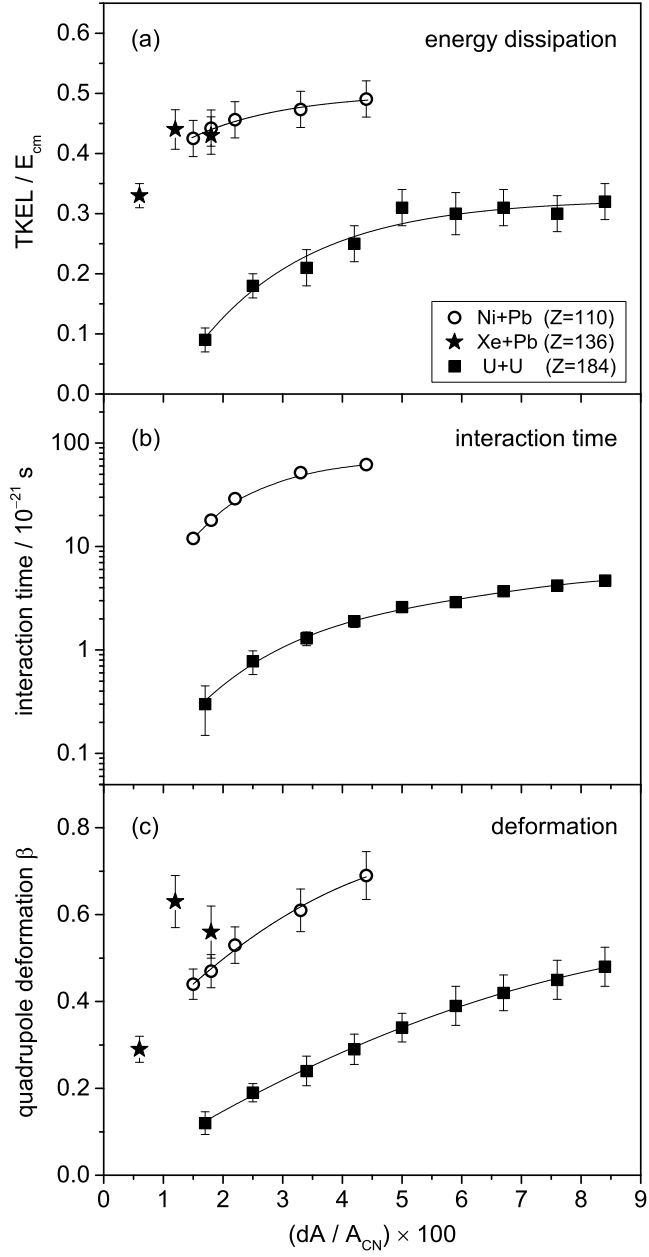}}
\vspace{-1mm}
\caption{
Measured behavior of total kinetic energy loss (TKEL) (a), interaction times (b)
and deformation of the fragments at the scission point (c) for collisions of
$^{64}$Ni+$^{207}$Pb ($E_{\rm c.m.}$\,=\,289~MeV),
$^{132}$Xe+$^{208}$Pb ($E_{\rm c.m.}$\,=\,492~MeV), and
$^{238}$U+$^{238}$U ($E_{\rm c.m.}$\,=\,875~MeV) as a function
of the fraction of transferred nucleons ${\rm d}A/A_{\rm CN}$.
}
\vspace{-3mm}
\label{FIG:EXPERIMENT}
\end{figure}

First, we show the experimental results for $^{64}$Ni+$^{207}$Pb ($E_{\rm c.m.}$\,=\,289~MeV),
$^{132}$Xe+$^{208}$Pb ($E_{\rm c.m.}$\,=\,492~MeV), and $^{238}$U+$^{238}$U
($E_{\rm c.m.}$\,=\,875~MeV) reactions (Fig.~\ref{FIG:EXPERIMENT}). The
lightest system, Ni$\,+\,$Pb, can still undergo fusion leading to isotopes of the
element darmstadtium~\cite{Hofmann(2001)}. Therefore, one can expect signatures
of formation of long-living nuclear molecules in this system. While the proton number of
the composite system U$\,+\,$U is far beyond the one of the heaviest known elements
and fusion reactions cannot be expected any more, it should give insight into the stability
of the heaviest accessible systems. To probe the stability and time evolution of the composite
nuclear systems, we investigate energy dissipation (\textit{i.e.}~TKEL), interaction
times, and quadrupole deformation of the exit channel nuclei at the scission point.
The deformation was extracted assuming that the binary reaction products have
the same quadrupole deformation and TKE is determined by the Coulomb potential
at the scission point~\cite{Heinz(2015)}.

In Fig.~\ref{FIG:EXPERIMENT}, we show experimentally measured TKEL (a),
interaction times (b), and quadrupole deformation (c) as a function of the fraction
of transferred nucleons ${\rm d}A/A_{\rm CN}$, where $A_{\rm CN}$ denotes
the total number of nucleons in the composite system. Figure~\ref{FIG:EXPERIMENT}
exhibits striking similarities between the different collision systems as well as between
the different parameters (TKEL, interaction time, and deformation). Two stages of
the reaction process are especially revealed by the behavior of TKEL and interaction
times. The first stage (${\rm d}A < 0.05 A_{\rm CN}$) is characterized by a steep
increase of these values, meaning that a large amount of energy is dissipated which
consumes a lot of time but only a small number of nucleons are transferred. This is the
transition regime from quasi-elastic to deep-inelastic reactions. After a net transfer
of about 5\% of the total number of nucleons in the composite system, the curves
turn and approach a saturation value. In this second stage, the situation reverses and
a large amount of nucleons can flow with small friction and small consumption of time.

It is striking that the slope change of TKEL and interaction times occurs, for so
different systems like Ni$\,+\,$Pb and U$\,+\,$U, always after the net transfer
of about 5\% of the total number of nucleons. Also the three TKEL values from
Xe$\,+\,$Pb indicate a similar trend. The evolution of the nuclear shapes shown
in Fig.~\ref{FIG:EXPERIMENT}~(c) exhibits a somewhat steeper increase of the
deformation at the beginning of the reaction, while in total the slope of the curves
is more uniform than in TKEL (a) and interaction times (b). This indicates that the
shape evolution proceeds more uniformly with increasing nucleon transfer than
TKEL and the interaction time.

\begin{figure}[t]
\centerline{%
\includegraphics[width=12.5cm]{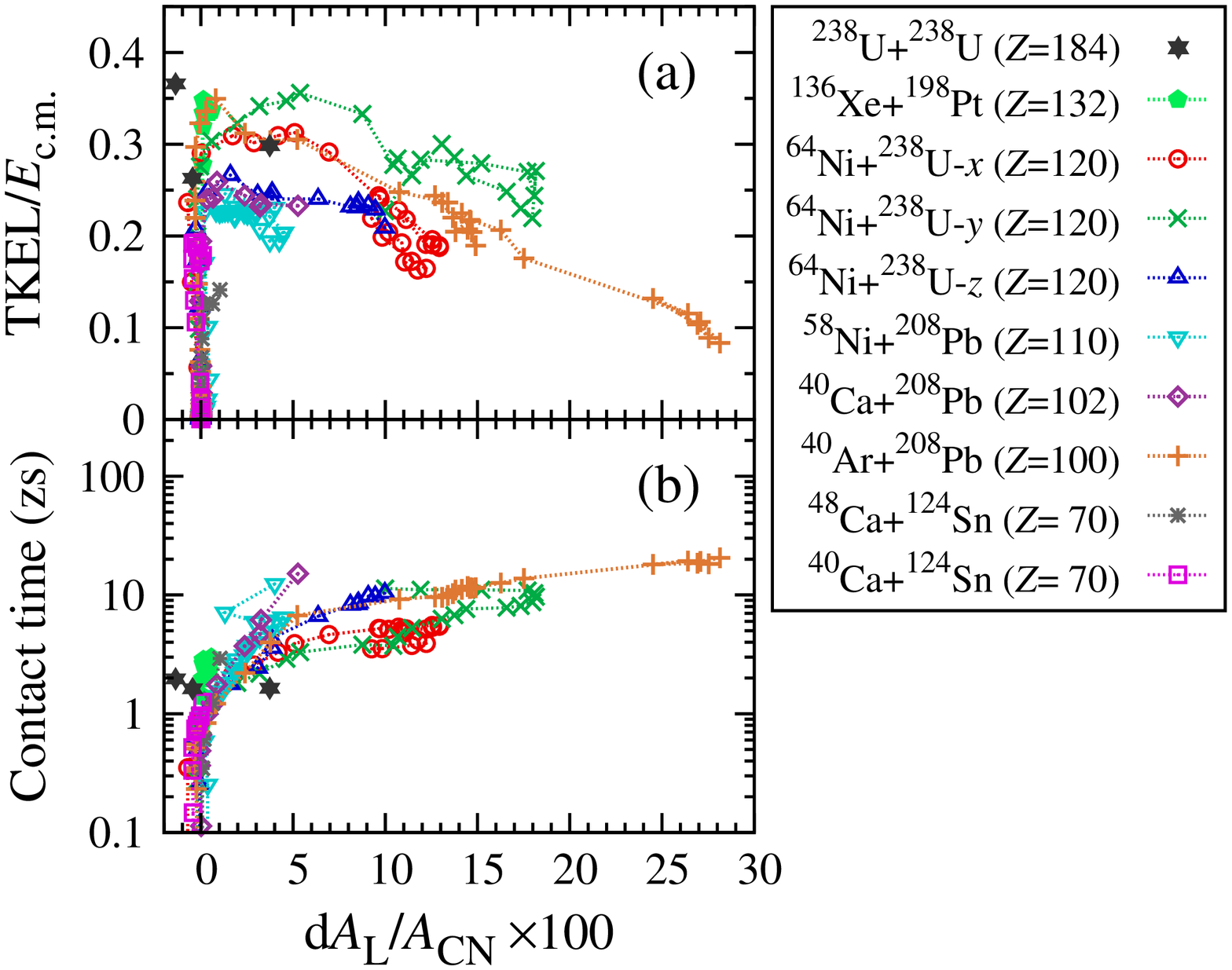}}
\caption{
Results of the TDHF calculations for various systems:
$^{40}$Ca+$^{124}$Sn (1.10),
$^{48}$Ca+$^{124}$Sn (1.09),
$^{40}$Ca+$^{208}$Pb (1.10),
$^{40}$Ar+$^{208}$Pb (1.37),
$^{58}$Ni+$^{208}$Pb (1.04),
$^{64}$Ni+$^{238}$U ($x$:~1.27, $y$ and $z$:~1.16),
$^{136}$Xe+$^{198}$Pt (1.19), and
$^{238}$U+$^{238}$U (tip-on-tip: 1.32, tip-on-side: 1.23, side-on-side: 1.13)
(The values in the parentheses are the ratio of center-of-mass energy
to the frozen HF barrier height, $E_{\rm c.m.}/V_{\rm B}$). Total kinetic
energy loss (TKEL) (a) and contact time (b) are shown. The horizontal axis
is the change in the mass number of the lighter nucleus relative to the total
number of nucleons in the composite system, ${\rm d}A_L/A_{\rm CN}$.
In panel (b), the contact time is shown in zeptosecond (1~zs\,=\,$10^{-21}$~sec).
}
\vspace{-3mm}
\label{FIG:TDHF}
\end{figure}

To understand the observed behavior, in the following, we investigate results
of TDHF calculations and make a possible comparison with the experimental
data. Nowadays it is feasible to systematically perform 3D TDHF calculations for
various projectile-target combinations, impact parameters, and incident energies
\cite{KS_KY_Ni-U,KS_KY_MNT,KS_KY_PNP,KS_KY_FUSION14,KS_KY_ARIS2014,
KS_KY_Maruhn,Bidyut(2015),KS_PhD}, using a parallel computational code which
works on hundreds of CPUs with MPI and OpenMP \cite{KS_PhD}. We show
results for various systems, namely,
$^{40}$Ca+$^{124}$Sn ($E_{\rm c.m.}$\,=\,129~MeV),
$^{48}$Ca+$^{124}$Sn ($E_{\rm c.m.}$\,=\,125~MeV),
$^{40}$Ca+$^{208}$Pb ($E_{\rm c.m.}$\,= \,209~MeV),
$^{58}$Ni+$^{208}$Pb ($E_{\rm c.m.}$\,=\,257~MeV) \cite{KS_KY_MNT},
$^{64}$Ni+$^{238}$U ($E_{\rm c.m.}$\,=\,307~MeV) \cite{KS_KY_Ni-U},
$^{136}$Xe+$^{198}$Pt ($E_{\rm c.m.}$\,=\,484~MeV) \cite{KS_PhD},
$^{40}$Ar+$^{208}$Pb ($E_{\rm c.m.}$\,=\,218~MeV), and
$^{238}$U+$^{238}$U ($E_{\rm c.m.}$\,=\,875~MeV). For all results presented
here, Skyrme SLy5 parameter set~\cite{Chabanat} was used for the EDF. TDHF
calculations were performed for various impact parameters, typically 0--10~fm, for
a given incident energy (except for $^{238}$U+$^{238}$U, see below). For the
$^{64}$Ni+$^{238}$U reaction, three initial orientations of deformed $^{238}$U
(prolate, $\beta$\,$\simeq$\,$0.27$) were investigated, where the symmetry axis
of $^{238}$U was set parallel to the collision axis ($x$~axis), impact parameter
vector ($y$~axis), or perpendicular to the reaction plane (parallel to $z$~axis)
\cite{KS_KY_Ni-U}. For axially symmetric nuclei with a relatively small deformation
[$^{40}$Ar (oblate, $\beta$\,$\simeq$\,0.13),
$^{58}$Ni (prolate, $\beta$\,$\simeq$\,0.11),
$^{64}$Ni (oblate, $\beta$\,$\simeq$\,0.12),
$^{124}$Sn (oblate, $\beta$\,$\simeq$\,0.11), and
$^{136}$Xe (oblate, $\beta$\,$\simeq$\,0.06)\footnote{The state
of $^{136}$Xe that was used for TDHF calculations turned out to be of a local
minimum with 40~keV higher energy than the HF ground state
(triaxial, $\beta$\,$\simeq$\,0.06 with $\gamma$\,$\simeq$\,29$^\circ$).
We note that no significant change is expected with this small deformation.}],
their symmetry axis was always set perpendicular to the reaction plane.
For $^{198}$Pt with a triaxial deformation ($\beta$\,$\simeq$\,0.12 with
$\gamma$\,$\simeq$\,33$^\circ$), the axis around which $|Q_{22}|$
takes the smallest value is set perpendicular to the reaction plane.
In reactions involving open shell nuclei, pairing correlations may play
an important role. We note, however, that we ignore the pairing effects
in the present article, as it requires additional computational effort.

First, we discuss the reaction mechanisms suggested by the TDHF calculations for
various systems. In Fig.~\ref{FIG:TDHF}, TKEL, divided by the center-of-mass energy
$E_{\rm c.m.}$, and contact time obtained from the TDHF calculations are shown in
panels (a) and (b), respectively. The horizontal axis is the change in the mass number of
the lighter nucleus, ${\rm d}A_L = \bigl<A_L\bigr>-A_L$, divided by $A_{\rm CN}$,
where $\bigl<A_L\bigr>$ is the average mass number of the lighter fragment. TKEL
was evaluated from the center-of-mass motion of the fragments~\cite{KS_KY_MNT}.
The contact time is defined as the time duration during which the lowest density between
two colliding nuclei exceeds a half of the nuclear saturation density, 0.08~fm$^{-3}$
\cite{KS_KY_Ni-U}.

Let us first focus on the cases where the average number of transferred nucleons is
small, a few percent of the total number of nucleons of the system. They correspond
to (quasi)elastic and grazing reactions. In Fig.~\ref{FIG:TDHF}, the TDHF results exhibit
a prominent increase of TKEL up to 20--35\% of $E_{\rm c.m.}$ and of contact time
up to about 1\,zs. This trend is common for all systems examined here. After the rapid
increase, TKEL is saturated, meaning that the available energy is fully transferred from
the relative motion to internal degrees of freedom. Note that for $^{40,48}$Ca+$^{124}$Sn
(purple squares and gray asterisks) with a relatively small charge product ($Z_{\rm P}
Z_{\rm T}$\,=\,1000) the results end before the TKEL saturation since the system
fused easily at smaller impact parameters. The TDHF results indicate that full energy
dissipation is quickly achieved at the first stage of the reaction.

Next let us look at the cases where the average number of transferred nucleons
is greater than a few percent of the total number of nucleons of the system. They
correspond to trajectories at smaller impact parameters. In such cases, a dinuclear
system connected with a thick neck is formed in the course of the collision and its shape
evolution dynamics is responsible for the amount of nucleon transfer. As the impact
parameter decreases, two nuclei collide more deeply, forming a more compact system
connected with a thicker neck. It makes the contact time longer and the system evolves
more towards the mass symmetry. It results in the behavior shown in Fig.~\ref{FIG:TDHF}~(b)
that indicates a correlation between the amount of mass transfer and the contact time.
In addition, Fig.~\ref{FIG:TDHF}~(a) shows a tendency that the TKEL decreases as
the number of transferred nucleons increases (${\rm d}A_L/A_{\rm CN}\gtrsim10$\%).
This is due to the fact that a larger mass transfer means also a larger proton transfer in
this QF regime. As the number of transferred protons increases, the charge product of
the fragments increases, which results in larger kinetic energy and, thus, smaller TKEL.
The TDHF results indicate that, in the second stage of the reaction, a large number of
nucleons are collectively transferred via shape (mean-field) evolution dynamics in the
composite system.

It is worth noting here similarities and differences of the TDHF results for different
systems. In Fig.~\ref{FIG:TDHF}~(a), it is shown that the saturated values of
TKEL for $^{40}$Ca+$^{208}$Pb, $^{58}$Ni+$^{208}$Pb, and $^{64}$Ni+$^{238}$U
($z$-direction, blue open triangles) are very similar to each other, about 25\%
of $E_{\rm c.m.}$. On the other hand, the TKEL saturates at larger values, about
30--35\% of $E_{\rm c.m.}$, for $^{40}$Ar+$^{208}$Pb, $^{136}$Xe+$^{198}$Pt,
and $^{64}$Ni+$^{238}$U ($x$- and $y$-direction, red circles and green crosses)
reactions. In the latter case, the significance of nuclear orientations has been pointed
out~\cite{KS_KY_Ni-U}. We note that the $y$-direction case (close to side collisions)
results in larger energy dissipation compared with the $x$-direction case (close to tip
collisions), for which one may expect the opposite trend since the barrier height is higher
in the side collisions. Significant roles of the shape evolution dynamics and shell effects
of $^{208}$Pb in those damped collisions were extensively discussed in Ref.~\cite{KS_KY_Ni-U}.
In the $^{40}$Ar+$^{208}$Pb and $^{136}$Xe+$^{198}$Pt reactions, the situation
is somewhat different. In this case, the collision energy is larger compared with the other
cases ($E_{\rm c.m.}/V_{\rm B}$\,$\simeq$\,1.37 and 1.19, respectively). Because
of this fact, a larger amount of energy is brought into the system that leads to a larger
maximum value of TKEL (note also that in the $^{136}$Xe+$^{198}$Pt reaction no
fusion reaction was observed and larger TKEL is achieved at smaller impact parameters).
Moreover, we observed several fusion-fission like processes in the $^{40}$Ar+$^{208}$Pb
reaction where the composite system splits in an almost symmetric way after a long contact
time ($\approx 20$~zs) which resulted in ${\rm d}A_L/A_{\rm CN} \simeq 28$ in
Fig.~\ref{FIG:TDHF}, because of the large angular momentum brought into the system.

Here, let us make a possible comparison between the TDHF results and the experimental
data. As mentioned above, the TDHF results correspond to contributions from main (most
probable) trajectories with various scattering angles associated with different impact parameters.
Whereas the experimental data shown in Fig.~\ref{FIG:EXPERIMENT} were obtained by
measurements for a fixed angular range: $\theta_{\rm lab}$\,=\,$(0\pm2$) degrees
(for Ni\,+\,Pb and Xe\,+\,Pb) and ($35 \pm 5$) degrees (for U\,+\,U). Therefore,
we should pay particular attention in comparing with the experimental data.

In the experiments on Ni\,+\,Pb \cite{Comas(2012)} and Xe\,+\,Pb \cite{Heinz(2015)}
at the velocity filter SHIP at GSI, only very central collisions were selectively detected, because
of the angular acceptance of $(0\pm2$) degrees. In central collisions, two nuclei must collide
deeply forming a dinuclear system. In such a case, full energy dissipation should be achieved
according to the TDHF results, as is also apparent from the long interaction time ($>10$~zs)
(cf.~Fig.~\ref{FIG:EXPERIMENT}~(b)). As shown in Fig.~\ref{FIG:EXPERIMENT}~(a),
the experimental data exhibit almost saturated values of TKEL, and thus, consistent with
the TDHF results. Comparing the TDHF results for $^{58}$Ni+$^{208}$Pb with the
experimental data for $^{64}$Ni+$^{207}$Pb, we find different TKEL values at saturation:
in the former case it is about 25\% of $E_{\rm c.m.}$, while in the latter case it is about
45--50\% of $E_{\rm c.m.}$. This difference should arise from the different beam energies.
In $^{58}$Ni+$^{208}$Pb the energy was $E_{\rm c.m.}/V_{\rm Bass}$\,$\simeq$\,0.97,
while in $^{64}$Ni+$^{207}$Pb it was $E_{\rm c.m.}/V_{\rm Bass}$\,$\simeq$\,1.11,
where $V_{\rm Bass}$ is the phenomenological fusion barrier \cite{Bass}. One may notice
here that, in the Xe\,+\,Pb case at ${\rm d}A/A_{\rm CN}$\,$<$\,1\% (Fig.~\ref{FIG:EXPERIMENT}~(a)),
a noticeably small value of TKEL was deduced. It may be related to shell effects of $^{208}$Pb
which hinder energy dissipation and also lead to a small deformation of the fragments, resulting
in larger TKE (smaller TKEL) \cite{KS_KY_Ni-U}. Although we have no experimental values for
Ni\,+\,Pb below ${\rm d}A/A_{\rm CN}$\,=\,1.5\%, one may expect the same behavior
also for Ni\,+\,Pb, extrapolating the curve down to ${\rm d}A/A_{\rm CN}$\,$<$\,1\%.
We note that it should also be influenced by the entrance-channel $N/Z$ asymmetry,
$|N_{\rm P}/Z_{\rm P} - N_{\rm T}/Z_{\rm T}|$, which is 0.24 for $^{64}$Ni+$^{207}$Pb
and 0.09 for $^{132}$Xe+$^{208}$Pb, \textit{i.e.}, in the former case it may weaken
the shell effects by the charge equilibration process \cite{Cedric(2012)}.

On the other hand, the experiment on U\,+\,U at the VAMOS spectrometer at GANIL
\cite{Golabek(2010)} was optimized for detecting fragments around the grazing angle,
which is relevant to the TDHF results for various systems shown in Fig.~\ref{FIG:TDHF}
(we expect that the universal behavior observed for other systems will also hold for U\,+\,U).
In Fig.~\ref{FIG:EXPERIMENT}~(a), the experimental data show a rapid increase of TKEL
for ${\rm d}A/A_{\rm CN}$\,$<$\,5\%, while it is almost saturated for larger mass transfers.
This behavior agrees with the reaction mechanisms deduced by TDHF, where substantial energy
dissipation takes place even with transfer of a small number of nucleons. Moreover, in Ref.~\cite{Golabek(2010)}
measurements were performed for several incident energies, and it was revealed that TKE
of the fragments becomes almost energy independent as the number of transferred nucleons
increases, ${\rm d}A/A_{\rm CN}$\,$\gtrsim$\,3\%. It is consistent with the observation
in TDHF where full energy dissipation is quickly achieved and a large number of nucleons are
transferred via the shape evolution dynamics in the composite system.

The experimental data shown in Fig.~\ref{FIG:EXPERIMENT}~(b) indicate that
a remarkable time delay (up to around $4$~zs) may still occur even for the heaviest
accessible system U\,+\,U. The interaction time was deduced from the measured
variance of the fragment masses $\sigma_A^2$ \cite{Golabek(2010)} assuming
the relation $\sigma_A^2 = 2D_A\tau$, given by a diffusion model \cite{Riedel(1979)},
where $D_A$ is the mass diffusion coefficient ($6.0 \times 10^{22}$~s$^{-1}$
for this reaction \cite{Golabek(2010)}) and $\tau$ is the interaction time. In order to
investigate reaction mechanisms further, we performed exploratory TDHF calculations
for head-on collisions of $^{238}$U+$^{238}$U at $E_{\rm c.m.}=875$~MeV with
three configurations, \textit{i.e.}, tip-on-tip, tip-on-side, and side-on-side collisions,
where in the side-on-side configuration the symmetry axes of two $^{238}$U were
aligned. The results are shown in Fig.~\ref{FIG:TDHF} by black stars.

In the experiment \cite{Golabek(2010)}, it has been clarified that more than 90\%
of the deep inelastic fragments resulted from $\tau$\,$<$\,2~zs. From the TDHF
calculations, we have obtained the contact times of 1.97~zs (tip-on-tip), 1.64~zs
(tip-on-side), and 1.62~zs (side-on-side), which are consistent with the experimental
observation. Although the contact time is rather short, substantial energy dissipation
takes place. The values of TKEL/$E_{\rm c.m.}$ are 0.37 (tip-on-tip), 0.30 (tip-on-side),
and 0.26 (side-on-side), which are again consistent with the experimental values shown
in Fig.~\ref{FIG:EXPERIMENT}~(a).

One may notice that, even in this symmetric system, reaction products can be different
from $^{238}$U. The reason is two-fold. One is due to the broken reflection symmetry
in the tip-on-side collision, which allows the system to split in an asymmetric way. In this
process, about 11.2 neutrons and 6.6 protons are transferred on average from tip-aligned
$^{238}$U to the other, consistent with earlier studies \cite{Cedric(2009),Kedziora(2010)}.
The other reason is due to a ternary QF process observed in the tip-on-tip collision. In the
latter case, an extremely long (more than 10~fm) neck is developed when the system evolves
towards the reseparation. The long neck becomes thinner at two points and eventually raptures
producing a small third fragment in between two heavy nuclei. We observed a beryllium-like
nucleus ($Z$\,$\simeq$\,4.1, $N$\,$\simeq$\,6.5) as the third fragment. The formation
of ternary fragments in tip-on-tip collisions of $^{238}$U+$^{238}$U was also reported
in Ref.~\cite{Cedric(2009)}. In this way, TDHF predicts significant impacts of nuclear
orientations on the fragment masses in collisions of two well-deformed actinide nuclei.

Finally, we investigate the variance of the fragment masses $\sigma_A^2$ in the
$^{238}$U+$^{238}$U reaction. Although TDHF provides a fairly good description for
averaged quantities, it substantially underestimates the variance of the fragment masses.
Thus one has to go beyond the standard mean-field description \cite{Lacroix,BV,SMF,TDDM}.
Here, we examine the variance employing the Balian-V\'en\'eroni (BV) variational principle
\cite{BV} which enables to include fluctuations and correlations around the mean-field
trajectory. For tip-on-side and side-on-side collisions, where we observed binary reaction
products\footnote{For tip-on-tip collision, where we observed the ternary QF process,
the BV prescription provided an unphysically large value of the variance (not shown).},
the BV prescription gives $\sigma_A^2$\,$\simeq$\,236.4 and 148.2, respectively, which
are significantly larger than those by TDHF, $\sigma_A^2$\,$\simeq$\,10.5 and 9.6.
For the tip-on-side and side-on-side collisions, we have TKEL values of 263 and 224~MeV,
respectively, and experimental values corresponding to those TKEL values are $\sigma_A^2
$\,$\approx$\,400 and 250 (cf.~Fig.~13 of Ref.~\cite{Golabek(2010)}). Although this
is a crude comparison, as calculations were performed only for head-on collisions and the
experimental variance is very sensitive to TKEL values, one can see that remarkable
improvement is achieved by the BV prescription. We note that the experimental data may
be influenced by the orientation dependence, as deduced by TDHF calculations, as well as
secondary processes (particle evaporation and fission), which may increase the variance
of the fragment masses, $\sigma_A^2$.

\section{Summary and perspective}\label{Sec:Summary}

We have performed theoretical and experimental studies on the stability
of heavy and superheavy nuclear systems with total proton numbers up to
$Z$\,=\,184 by investigating nuclear molecule formation and time delays
in deep-inelastic binary reactions at energies around the Coulomb barrier.

The experimental data for $^{64}$Ni+$^{207}$Pb ($E_{\rm c.m.}$\,=\,289~MeV)
\cite{Comas(2012)}, $^{132}$Xe+$^{208}$Pb ($E_{\rm c.m.}$\,=\,492~MeV)
\cite{Heinz(2015)}, and $^{238}$U+$^{238}$U ($E_{\rm c.m.}$\,=\,875~MeV)
\cite{Golabek(2010)} show striking similarities in the behavior of energy dissipation,
interaction times, and deformation of the fragment nuclei. In the first stage of the
reaction, where the amount of nucleon transfer is less than 5\% of the total number
of nucleons in the composite system, a lot of time is spent to move only a small number
of nucleons and large energy is transferred into internal excitations. In the second stage,
the interaction time increases slowly even if a large number of nucleons are transferred.
The observed similarities indicate that a significant time delay may still occur even in
the heaviest accessible system U$\,+\,$U.

To understand the observed behavior, we have carried out a comparative
study between results of TDHF calculations for various systems
[$^{40}$Ca+$^{124}$Sn ($E_{\rm c.m.}$\,=\,129~MeV),
$^{48}$Ca+$^{124}$Sn ($E_{\rm c.m.}$\,=\,125~MeV),
$^{40}$Ca+$^{208}$Pb ($E_{\rm c.m.}$\,= \,209~MeV),
$^{40}$Ar+$^{208}$Pb ($E_{\rm c.m.}$\,=\,218~MeV),
$^{58}$Ni+$^{208}$Pb ($E_{\rm c.m.}$\,=\,257~MeV),
$^{64}$Ni+$^{238}$U ($E_{\rm c.m.}$\,=\,307~MeV),
$^{136}$Xe+$^{198}$Pt ($E_{\rm c.m.}$\,=\,484~MeV), and
$^{238}$U+ $^{238}$U ($E_{\rm c.m.}$\,=\,875~MeV)]
and the experimental data. From the results of the TDHF calculations for different
systems, we have found similar trends as observed in the experimental data.

The TDHF results have revealed occurrence of two distinct transfer mechanisms.
In the grazing regime, a small number of nucleons are transferred through a fast
charge equilibration process. At this stage substantial energy dissipation (TKEL/$
E_{\rm c.m.}$\,$\approx$\,20--35\%) takes place in a relatively short period
(about 1--2~zs) and available kinetic energy is fully dissipated. This represents
a rapid transition from quasi-elastic to deep-inelastic and QF regimes. As the
impact parameter decreases, two nuclei collide more deeply forming a dinuclear
system connected with a thick neck. In the latter case shape evolution dynamics
of the composite system is responsible for the amount of nucleon transfer. A large
number of nucleons are effectively transferred via the shape evolution dynamics,
while all kinetic energy is already fully dissipated.

TDHF calculations were also performed for head-on collisions of $^{238}$U+ $^{238}$U
($E_{\rm c.m.}$\,=\,875~MeV) with three orientations, \textit{i.e.}, tip-on-tip,
tip-on-side, and side-on-side collisions. The results of TKEL and contact time are in
reasonable agreement with the experimental data. In the tip-on-tip collision, we
have observed a ternary QF process, where a small third fragment is generated from
the neck region between two heavy nuclei. In the tip-on-side collision, the broken
reflection symmetry allows the composite system to split in an asymmetric way,
resulting in transfer of many nucleons from tip-aligned $^{238}$U to the other.
Moreover, the Balian-V\'en\'eroni variational principle was applied to investigate
the variance of the fragment masses in the U\,+\,U system, showing significant
improvement compared with TDHF.

A careful observation of the experimental data revealed striking similarities between
Ni$\,+\,$Pb and U$\,+\,$U systems that the slope change of TKEL and interaction
times occurs always after the net transfer of about 5\% of the total number of nucleons.
We could not yet draw a conclusion if a physical meaning is behind or it is just by chance.
We consider that further analyses of different reactions measured at SHIP or VAMOS
will provide useful information, which we leave as a future task.

It has been shown that the microscopic TDHF theory can be a promising tool to investigate
the QF dynamics in heavy and superheavy systems. The theory provides valuable insight
into the complex QF dynamics from nucleonic degrees of freedom, thus, taking into account
shell effects during the collision process. Recently, it has been suggested, based on the
Langevin model, that multinucleon transfer and QF processes are useful to produce neutron-rich
(super)heavy nuclei which have not been produced to date, where shell effects are predicted
to play an important role (see, \textit{e.g.}, \cite{Langevin:160Gd+186W}). In principle,
TDHF can also describe the predicted shell-effect driven transfer processes, and an extensive
analysis is in progress \cite{KS(IQF)}.

Last but not least, there is an open problem that how and to what extent the pairing
correlations affect the reaction dynamics in damped collisions of heavy nuclei. The so-called
time-dependent superfluid local density approximation (TDSLDA) (or time-dependent
Hartree-Fock-Bogoliubov theory, TDHFB) would provide a satisfactory answer to it,
although it requires about 100--1000 times larger computational cost compared with
that for the TDHF calculation. Very recently, it has become possible to perform 3D TDSLDA
calculations for nuclear systems using leadership-class supercomputers with hundreds of
GPUs \cite{CoulEX,PM,Fission,VortexPRL,VortexNIC}. On one hand, pairing effects should
disappear as the excitation energy increases, on the other hand the additional degrees
of freedom associated with complex pairing field dynamics may still alter the QF timescale
in a similar way as observed in $^{240}$Pu induced fission process \cite{Fission}.
A study along this line is in progress \cite{PGK}, and the results will be published elsewhere.

\section*{Acknowledgments}
K.S.~acknowledges support of Polish National Science Centre (NCN) Grant,
decision No.~DEC-2013/08/A/ST3/00708. The numerical calculations were performed
using computational resources of the HPCI system (HITACHI SR16000/M1) provided by
Information Initiative Center, Hokkaido University, through the HPCI System Research
Projects (Project IDs: hp120204, hp140010, hp150081, and hp160062), and using
computational resources of the Cray XC40 supercomputer system at the Yukawa
Institute for Theoretical Physics (YITP) at Kyoto University.

\end{document}